# Design of a stiffness adjustable magnetic fluid shock absorber based on optimal stiffness coefficient

Yanwen Li[1], Decai Li [1*], Yingsong Li[2]

[1]State Key Laboratory of Tribology, Department of Mechanical Engineering, Tsinghua University, Beijing, China

[2]China Productivity Center for Machinery Co., Ltd., Beijing, China

**\* Correspondence:**
Corresponding Author
lidecai@mail.tsinghua.edu.cn

**Abstract**

With the rapid development of aerospace technology, the vibration problem of the spacecraft flexible structure urgently needs to be solved. Magnetic fluids are a type of multi-functional smart materials, which can be employed in shock absorbers to eliminate these vibrations. Referring to the calculation methods of stiffness coefficients of other passive dampers, the stiffness coefficient formula of magnetic fluid shock absorbers (MFSAs) was derived. Meanwhile, a novel stiffness adjustable magnetic fluid shock absorber (SA-MFSA) was proposed. On the basis of the second-order buoyancy principle, a series of SA-MFSAs were fabricated. The range of stiffness coefficients covered by these SA-MFSAs contains the optimal stiffness coefficient estimated by formulas. The repulsive force measurement and vibration attenuation experiments were conducted on these SA-MFSAs. In the case of small amplitude, the relationship between the repulsive force and the offset distance was linear. The simulation and experiment curves of repulsive forces were in good agreement. The results of vibration attenuation experiments demonstrated that the rod length and the magnetic fluid mass influence the damping efficiency of SA-MFSAs. In addition, these results verified that the SA-MFSA with the optimal stiffness coefficient performed best. Therefore, the stiffness coefficient formula can guide the design of MFSAs.

## 1   Introduction

The vibration problem of the spacecraft's flexible solar panel is one of the most critical issues for the normal operation of spacecrafts [1]. The solar panels are susceptible to residual oscillation and driving disturbance [2], because of their characteristics of small damping and low frequency, which results in vibration. These micro-vibrations are difficult to reduce in the space environment so that vibration suppression of spacecraft's flexible structures has received a lot of attention in the past two decades [3].

One promising solution for this problem is to utilize shock absorbers with smart materials. Magnetic fluids (MFs) are a kind of smart materials, which are composed of nanoparticles, carrier liquids and surfactants [4]. Due to their chemical composition, MFs have a lot of good characteristics, such as rapid magnetic response, complex rheology, amazing levitation and so on [5]. According to these interesting properties, magnetic fluid shock absorber (MFSA) has many advantages, for instance, controllable damping, compact structure, high sensitivity, less energy consuming and long life, etc. [6] These good features make MFSAs suitable for vibration suppression with low frequency and small amplitude.

The first successful application of the MF viscous damper was presented by NASA in 1967, which was employed to suppress the oscillations of the Radio Astronomy Explorer Satellite in aerospace [7]. Afterwards, MF dampers attracted more and more attention and scholars from various countries devoted great efforts to the development of MF dampers [8]. Ten years later, on the basis of levitation characteristics of MFs, Moskowitz et al. [9] proposed a viscous fluid inertia damper working well in reducing the rotational vibrations of stepping motor shafts. In 2002, Bashtovoi et al. [10] created a novel MF dynamic absorber, which was regarded as a major breakthrough for dampening the spacecraft vibration. In order to improve its sensitivity, this MF dynamic absorber wasn't filled with MFs, which played an important role in subsequent development of MF inertia dampers. In the same period, on account of the controllable flow of MFs, researchers replaced ordinary liquids in tuned liquid dampers with MFs to raise their damping efficiency, which were called tuned MF dampers [11-13]. Due to the aggravation of energy problems, energy harvesting has received increasing attention. Vibration energy harvesters have become a hotspot for studies on MF dampers in the past decade [14, 15]. They are combination of energy harversters and MF dampers and can realize self-energizing, which boosts the development of equipment miniaturization [16]. In the previous extensive literature, the damping coefficient of MF dampers has always been concerned [17]. However, there are far fewer studies involving stiffness coefficient.

Fortunately, there are sufficient studies on the stiffness coefficients of other dampers. The research approaches about stiffness coefficients in these studies can be transferred to MF dampers. Sun et al. [18] presented a novel compact shock absober with variable stiffness, which was suitable for vehicle suspension. Its stiffness was controlled by current applied to the shock absorber. Wang et al. [19] proposed a new tuned inerter negative stiffness damper for protecting primary structures under earthquake excitaions. Its negative stiffness combined with inertance were employed to enhance the energy dissipation ratio. Javanbakht et al. [20] developed an analytical model to refine damper design by considering the negative and positive stiffness. The important parameters of negative stiffness dampers and positive stiffness dampers were predicted by the refined design formula. These design tools made dampers more sutiable to mitigate vibrations of stay cables. Weber et al. [21] created a novel adaptive tuned mass damper, of which damping and stiffness can be adjusted. This damper contained a magnetorheological damper that was used to control friction-viscous damping and stiffness of the whole damper. Combined with a controllable magnetorheological damper, this presented damper performed well over a wide frequency range.

In the paper, the optimal stiffness coefficient formula was derived in detail and a novel MF damper with adjustable stiffness was presented. In section 2, the whole derivation process of the optimal stiffness coefficient was elaborated. Then, the design method of a new stiffness adjustable MF shock absorber (SA-MFSA) was proposed in section 3. Section 4 illustrated the experimental process and corresponding apparatus. The simulation and experiment results of repulsive force, as well as the vibration attenuation experiment results were given and discussed in section 5. Finally, the main conclusions were outlined in section 6.

## 2 Theoretical analysis

### 2.1 Oscillation model

The cantilevered elastic plate represented the simplified flexible solar panel of spacecrafts and generated maximum vibration at the end. Therefore, SA-MFSAs were generally placed at the end of plate, as shown in Fig. 1a. Ulteriorly, the corresponding physical model was extracted from the experimental system model in Fig. 1a, which was called two-degree-of-freedom oscillation model, as shown in Fig. 1b.

Based on the vibration theory, the motion equations of the oscillation system in Fig. 1b are:

$$m_1 \ddot{x}_1 + C_1 \dot{x}_1 + C_2 (\dot{x}_1 - \dot{x}_2) + K_1 x_1 + K_2 (x_1 - x_2) = F$$

$$m_2 \ddot{x}_2 - C_2 (\dot{x}_1 - \dot{x}_2) - K_2 (x_1 - x_2) = 0$$

(1)

Where $x_1$ and $x_2$ are the displacement of the brass plate and the working unit respectively, $m_1$ is the equivalent mass of the brass plate and housing, $m_2$ is the equivalent mass of the working unit, $K_1$ is the equivalent stiffness coefficient of the brass plate, $K_2$ is the equivalent stiffness coefficient of the SA-MFSA, $C_1$ is the equivalent damping coefficient of the brass plate, $C_2$ is the equivalent damping coefficient of the SA-MFSA and $F$ is the sinusoidal excitation force. In this paper, the main parameter we focused on was the equivalent stiffness coefficient of the SA-MFSA, of which the symbol was $K_2$.

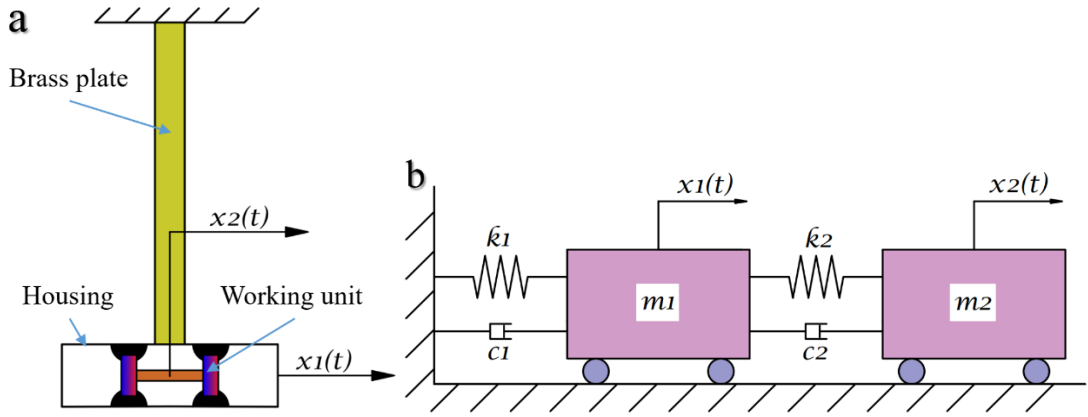

Figure 1. Simplified model. (a) Model of the experimental system. (b) Model of the oscillation system.

According to the continuous beam theory, the equivalent stiffness coefficient of the brass plate can be expressed as,

$$K_1 = \frac{3EI}{l_c^3} \tag{2}$$

Where $E = 9.7 \times 10^{10} Pa$ is the elastic modulus of the brass, $I = ab^3/12$ is the cross-section inertia of the brass plate, $l_c = 1.1m$, $a = 0.05m$ and $b = 0.005m$ are the length, width and thickness of the brass plate, respectively. From equation (2), $K_1 = 113.87 N/m$ is obtained.

On the basis of the calculation method of the cantilever beam with a lumped mass at one end, the first-order natural frequency of the brass plate is,

$$f = \frac{1}{2\pi}\sqrt{\frac{K_1}{m_1}} \tag{3}$$

$$m_1 = m_0 + \frac{33}{140} m_c$$

Where $m_0 = 1.55 kg$ and $m_c = 2.25 kg$ are the lumped mass and the own mass of the brass plate, respectively, which acquires that $m_1 = 2.08 kg$. Due to obtained $K_1$ and $m_1$, it is obvious that $f = 1.18 Hz$.

In order to facilitate the calculation, the sinusoidal excitation force and displacements are expressed in the form of complex variables function respectively.

$$F = \bar{F} e^{jwt} \tag{4}$$

$$x_1 = \overline{X_1} e^{jwt}$$

$$x_2 = \overline{X_2} e^{jwt}$$

By solving Equation (1) and applying $\overline{X_1}/\overline{F}$, $\overline{X_2}/\overline{F}$ as description form, that is,

$$\frac{\overline{X_1}}{\overline{F}}(w) = \frac{-m_2 w^2 + K_2 + jC_2 w}{\Delta_1}$$

$$\frac{\overline{X_2}}{\overline{F}}(w) = \frac{K_2 + jC_2 w}{\Delta_1} \tag{5}$$

$$\Delta_1 = [m_1 m_2 w^4 - (m_1 K_2 + m_2 K_1 + m_2 K_2 + C_1 C_2) w^2 + K_1 K_2] + j\{(C_2 K_1 + C_1 K_2)w - [m_1 C_2 + m_2(C_1 + C_2)]w^3\}$$

By adopting euler formula, the amplitude in the form of complex variables is subsitituted with the real amplitude and phase.

$$\psi_1 = \frac{X_1}{X_{st}}(w) = \frac{\sqrt{(\lambda^2 - \gamma^2)^2 + 4\lambda^2 \varsigma^2 \gamma^2}}{\sqrt{\Delta_2}}$$

$$\psi_2 = \frac{X_2}{X_{st}}(w) = \frac{\sqrt{\lambda^4 + 4\lambda^2 \varsigma^2 \gamma^2}}{\sqrt{\Delta_2}} \tag{6}$$

$$\Delta_2 = 4\lambda^2 \varsigma^2 \gamma^2 \{4\zeta^2 \gamma^2 + [1 - (1+\mu)\gamma^2]^2\} + 8\lambda \zeta \gamma^6 \mu \varsigma + \{\gamma^4 - [1 + (1+\mu)\lambda^2]\gamma^2 + \lambda^2\}^2 + 4\zeta^2 \gamma^2 (\lambda^2 - \gamma^2)^2$$

Where $X_{st} = \overline{F}/K_1$ (m) is the static deformation of the main system, $\varsigma = C_2/(2\sqrt{m_2 K_2})$ is the damping ratio of the SA-MFSA, $\zeta = C_1/(2\sqrt{m_1 K_1})$ is the damping ratio of the main system, $\mu = m_2/m_1$ is the mass ratio, $\lambda = w_n/\Omega_n$ is the natural angular frequency ratio, $\gamma = w/\Omega_n$ is the forced vibration frequency ratio, $w_n = \sqrt{K_2/m_2}$ (rad/s) is the natural angular frequency of the SA-MFSA, $\Omega_n = \sqrt{K_1/m_1}$ (rad/s) is the natural angular frequency of the main system.

By using Equation (6) which determains the amplitude ratio of the whole oscillation system, one of the most important parameters of SA-MFSAs, namely $K_2$, can be optimized design.

## 2.2 Optimal stiffness coefficient

In the case of ignoring the damping of the main system, which means that $C_1 = 0$, the optimal stiffness coefficient of the SA-MFSA can be calculated by,

$$K'_{2\_opt} = m_2 \frac{K_1}{m_1} \lambda_{opt}^2 = m_2 \frac{K_1}{m_1} \left(\frac{1}{1+\mu}\right)^2 \tag{7}$$

Where $\lambda_{opt}$ represents the optimal value of the natural angular frequency ratio.

When the mass ratio $\mu = 0.1 \sim 0.3$ and the damping ratio $\zeta = 0 \sim 0.125$ are within the specified range, through the numerical calculation, the optimal natural angular frequency ratio is refined as,

$$\lambda_{opt} = \frac{1}{1+\mu} - (0.241 + 1.74\mu - 2.6\mu^2)\zeta - (1 - 1.9\mu + \mu^2)\zeta^2 \tag{8}$$

By considering the damping of the main system, which means that $C_1 \neq 0$, the optimal stiffness coefficient of the SA-MFSA can be expressed as,

$$K_{2opt} = m_2 \frac{K_1}{m_1} \lambda_{opt}^2$$

$$= m_2 \frac{K_1}{m_1} \left[\frac{1}{1+\mu} - (0.241 + 1.74\mu - 2.6\mu^2)\zeta - (1 - 1.9\mu + \mu^2)\zeta^2\right]^2 \tag{9}$$

By putting into specific value, the optimal stiffness coefficient of the SA-MFSA was got, that is $K_{2opt} = 5.25 N/m$.

## 3    Structure design

Fig. 2 depicts the section view of the SA-MFSA, including the major components and main dimensions. Meanwhile, Table 1 lists the main structural parameters, corresponding symbols and values. The connecting rod linked two moving magnets at both ends and the moving magnets adsorbed the MF, which formed the working unit. The housing adopted the split strucutre and was fabricated with holes using to balance the pressure in the cavity. The static magnet was fastened with two fixing spacers and aligned with the moving magnets to provide the restoring force. The rubber rings were mounted on the fixing spacers to ensure the MF was sealed. Fig. 2 also shows the materials of each components. It's worth to notice that the initial connecting rod was made of copper, but the change length of the connecting rod used resin to aviod the influence of the weight alteration. The whole appearance of the SA-MFSA is shown in Fig. 3.

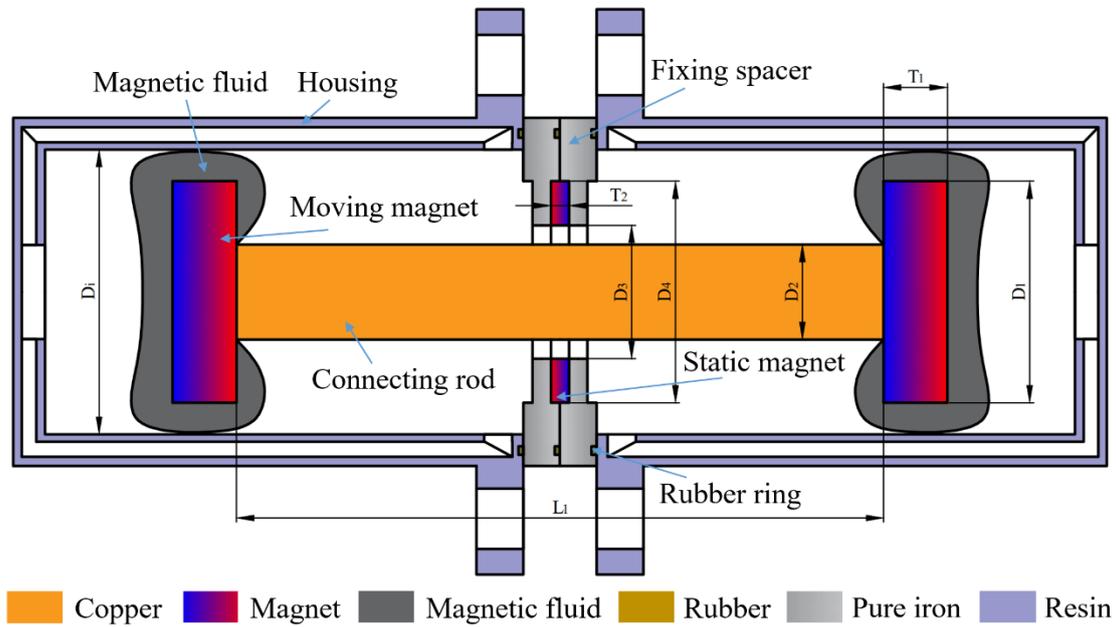

Figure 2. Section view of the stiffness adjustable magnetic fluid shock absorber.

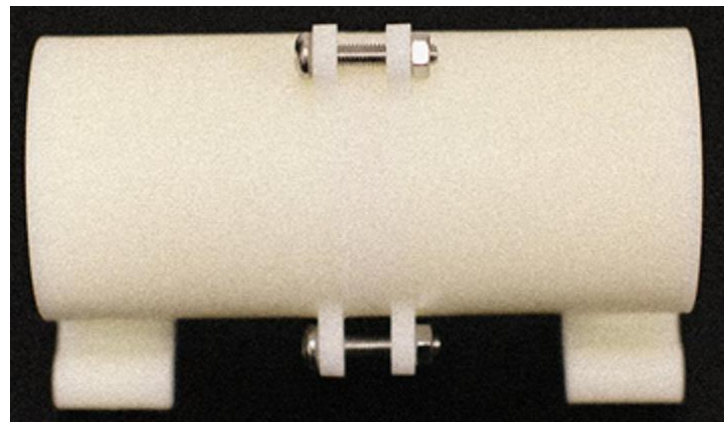

Figure 3. Appearance view of the stiffness adjustable magnetic fluid shock absorber.

Table 1. Main structural dimensions of the stiffness adjustable magnetic fluid shock absorber.

| Parameters | Symbols | Values |
| --- | --- | --- |
| Diameter of the moving magnet | $D_1$ | 30 mm |
| Thickness of the moving magnet | $T_1$ | 5 mm |
| Diameter of the connecting rod | $D_2$ | 10 mm |
| Initial length of the connecting rod | $L_1$ | 88 mm |
| Inner diameter of the static magnet | $D_3$ | 30 mm |
| Outer diameter of the static magnet | $D_4$ | 20 mm |
| Thickness of the static magnet | $T_2$ | 2 mm |
| Inner diameter of the housing | $D_i$ | 36 mm |

## 4  Experiments

It can be seen from Fig. 4 that the vibration system was composed of a vibration table and a data acquisition system. The vibration table involved a base and a holder to immobilize one end of the brass plate. The other end of the brass plate was fitted with a frame using to install the SA-MFSA. The non-magnetic support stretched out from the holder for fixing the laser displacement sensor. The size of the cantilever elastic brass plate was 1100mm*50mm*5mm, as well as the small amplitude and low frequency vibrations were generated by the plate. The frequency of the free oscillations of the cantilever elastic brass plate was equal to 1.18Hz and the amplitude was set to 6mm in the subsequent vibration reduction experiments. As for the data acquisition system, it included a acquisition card, a computer and an laser displacement sensor. The acquisition card USB-DAQ-7606i collected electrical signals and transmitted them to the computer with sampling frequency of 50Hz. In order to meet the requirement of experiments, the laser displacement sensor HL-G108-A-C5 was chosen, of which the resolution was 2.5um.

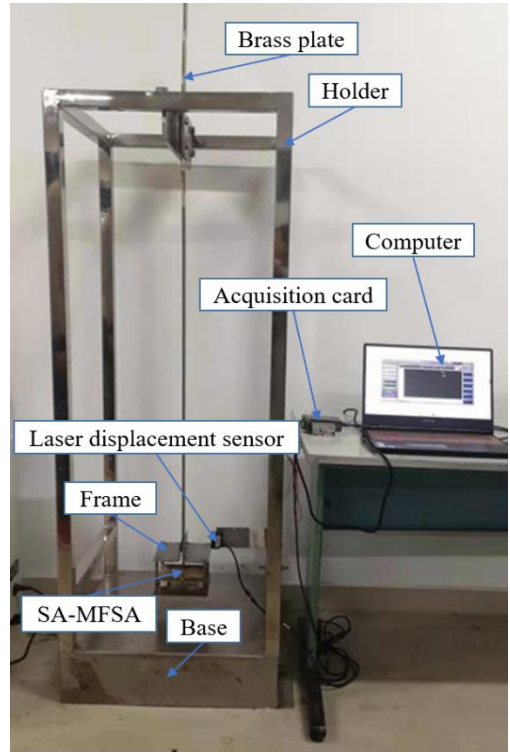

Figure 4. Photograph of experimental apparatus of vibration system.

The experiment apparatus shown in Fig. 5 was employed to measure the repulsive force between the static magnet and moving magnets. The dynamometer was connected with the sliding table through the fixing plate. The eletronic slide gauge was concatenated with the sliding table to obtain the position of the dynamometer. The accuracies of the dynamometer and the eletronic slide gauge were 0.001N and 0.01mm, respectively. The static magnet was fixed on the base made of resin, as well as the moving magnet was attached to the non-magnetic rod, which ensured the results weren't affected by extra magnetic field. Besides, the non-magnetic rod, moving magnets and the static magnet were coaxial. First, we made the connecting rod center coincide with the static magnet center by adjusting the lifting platform and the sliding table, then set the dynamometer to zero. Second, rotating the rocking handle let moving magnets leave the equilibrium position and controlled the movement step length at 0.2mm. In the end, the offset distances and corresponding repulsive forces were transmitted to the computer for subsequent analysis.

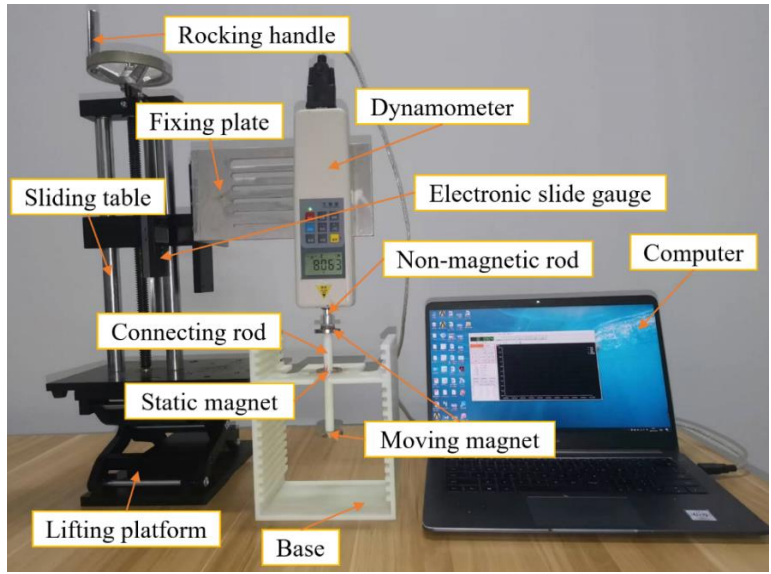

Figure 5. Photogragh of experimental apparatus to measure repulsive force.

The MF used in experiments was prepared by our laboratory, of which the density and viscosity were 1.23g/cm$^3$ and 0.26Pa s, respectively. Fig. 6 gave the magnetization curve of the MF.

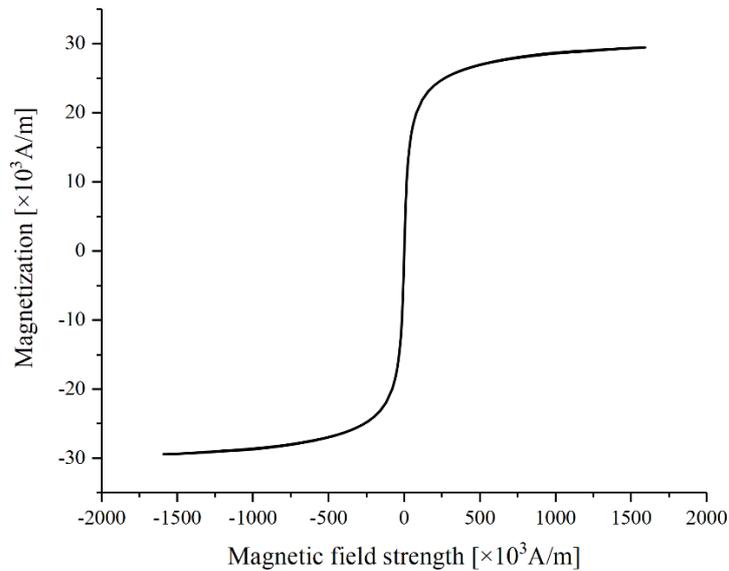

Figure 6. Magnetization curve of the magnetic fluid used in experiments.

## 5   Results and discussion

### 5.1   Stiffness coefficient

The length range of the connecting rod was 88mm to 112mm and the increment between adjacent rods was 4mm. The static magnet and the working unit with adjustable connecting rod, of which dimensions were same as Table 1, were put into COMSOL for simulation. As a result, the relationship between the repulsive force and the offset distance was obtained and shown in Fig. 7. The experimental curves described the relationship between the repulsive force and the offset distance in Fig. 8, which were measured by experimental apparatus in Fig. 5. The offset distance was adjusted from 0mm to 12mm and the interval was 0.1mm. Since the minimum reading of the dynamometer was 0.02N, all experimental curves started from the repulsive force equal to 0.02N. The maximum amplitude of vibration reduction experiments was only 6mm, so actual offset distance wouldn't exceed 6mm. In this case, the repulsive force is linear with the offset distance, as shown in Fig. 7 and Fig. 8. Therefore, the equation of stiffness coefficient can be approximated as:

$$F_m \approx K_2 x \qquad (10)$$

Where $F_m$ is the repulsive force and $x$ is the offset distance.

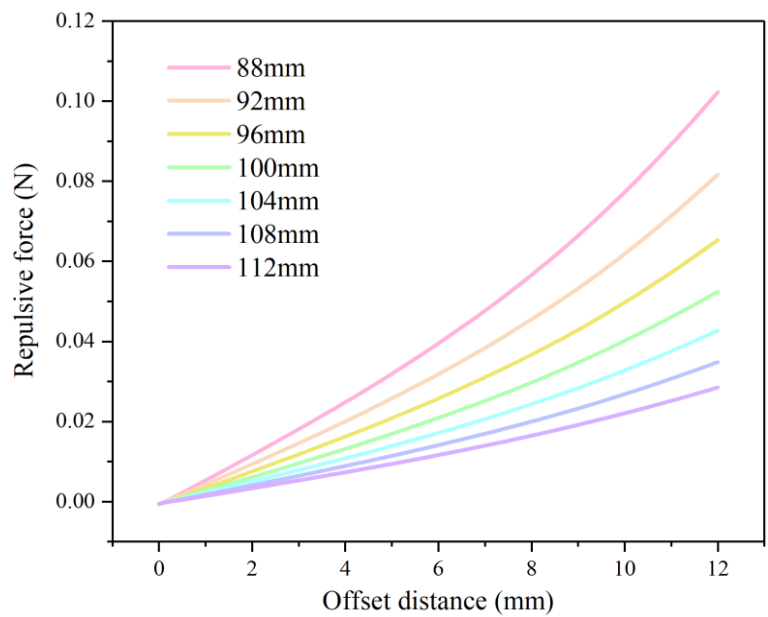

Figure 7. Simulated repulsive forces for different offset distances with various rod lengths.

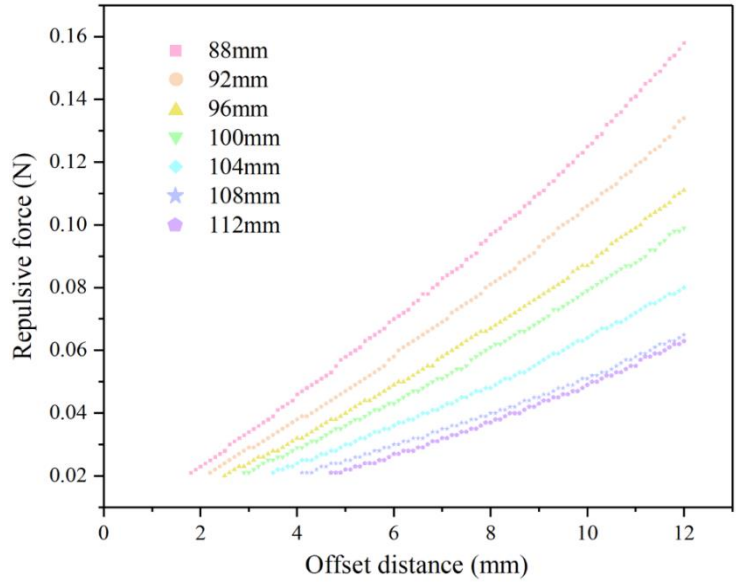

Figure 8. Measured repulsive forces for different offset distances with various rod lengths.

The simulation and experiment results of the stiffness coefficient with different rod lengths were calculated separately by equation (10). When rod length was equal to 108mm, the simulation and experiment resluts of the stiffness coefficinet were 5.05N/m and 5.15N/m, respectively. Both results were close to the calculation result 5.25N/m. At the same time, it can be seen from Fig. 9 that the simualtion and experiment results were in good aggrement, especially when the rod lengths were taken from 96mm to 108mm. This illustrated that it was feasible to design SA-MFSA structure according to the optimal stiffness coefficient. The specific operation steps were: (1) First, through the theoretical model, the optimal stiffness coefficient was estimated. (2) Second, the SA-MFSA structure was designed based on simulation, which meant that adjusting the rod length made the optimal stiffness coefficient included within the rod length range. (3) Finally, the corresponding SA-MFSAs were fabricated and the correctness was verified by experimental apparatus in Fig. 5.

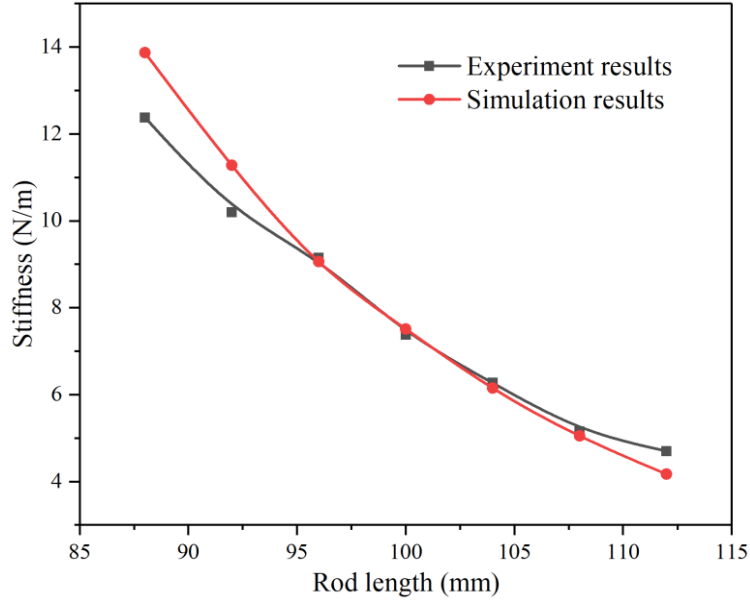

Figure 9. Compared simulation and experiment results of stiffness.

**5.2 Vibration reduction**

The results of the vibration attenuation experiments were described in terms of the attenuation time and the attenuation amplitude, respectively. From the perspective of attenuation time, vibration attenuation rate $\eta$ was introduced to quantify the damping efficiency of SA-MFSAs, of which the definition is as follow.

$$\eta = 1 - \frac{T_n}{T_0} \tag{11}$$

Where $T_0$ is the time taken by the equivalent mass when the amplitude decays to 2% of the initial amplitude, $T_n$ is the time taken by the SA-MFSA when the amplitude decays to 2% of the initial amplitude.

Fig. 10 depicts the relationship between vibration attenuation rate and rod length under different amount of MF. It is obvious that vibration attenuation rate increased with rod length becoming longer. In addition, it reached maximum value as rod length taken 108mm and would slightly decrease when rod length exceeded 108mm. Regarding the amount of MF, the following two points were mainly considered. First, in order to ensure the stable suspension of the working unit, MF mass should be greater than minimum critical value. Second, if using too much MF, the moving magnet couldn't fully adsorb MF and the extra MF might be bad for the performance of SA-MFSAs. Therefore, we chose 7g to 15g MF on each moving magnet for experiments. It is observed that when MF ranged from 7g to 11g, vibration attenuation rate continuously enhanced, which meant energy consumption became faster with more MF. However, when MF exceeded 11g, vibration attenuation rate began to decrease. This was because too much MF hindered the working unit movement, which was detrimental to the damping performance.

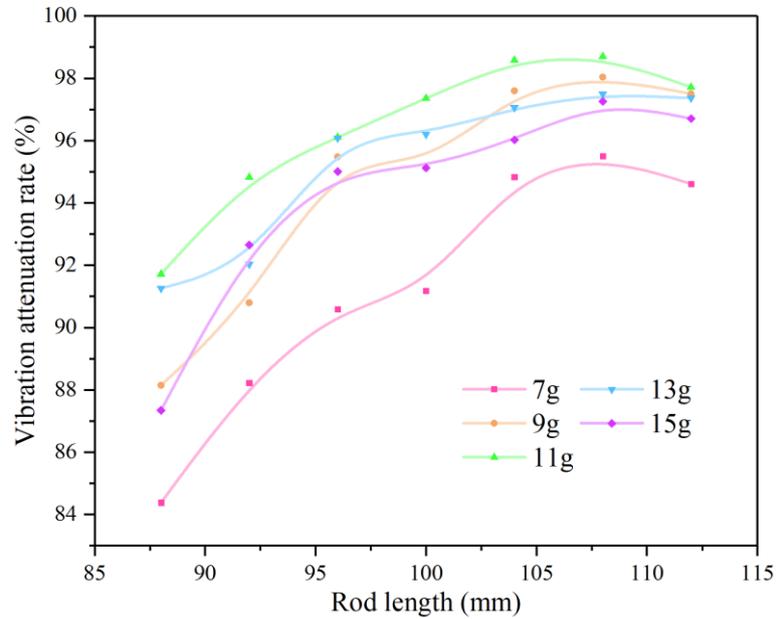

Figure 10. Vibration attenuation rate versus rod length with different magnetic fluid mass.

According to the attenuation amplitude, logarithmic decay rate, another dimensionless number was used to quantify the damping efficiency of SA-MFSAs, of which the calculation equation is as below.

$$\delta = \frac{1}{i} ln \left(\frac{A_0}{A_i}\right) \qquad (12)$$

Where $A_0$ is the initial amplitude, $A_i$ is the amplitude of $i$th oscillation and $i$ is equal to 5.

Fig. 11 represents the dependence of logarithmic decay rate on rod length with different MF mass. Just like vibration attenuation rate, logarithmic decay rate also increased with rod lengthened. In the rod length range of 100mm to 108mm, logarithmic decay rate grew rapidly, but it dropped slowly as rod length longer than 108mm. The peak of logarithmic decay rate also occurred at rod length equal to 108mm. As for the effect of MF mass on logarithmic decay rate was basically the same as that on vibration attenuation rate. In general, both vibration attenuation rate curve and logarithmic decay rate curve proved that the SA-MFSA presented the best damping performance with rod length taken 108mm.

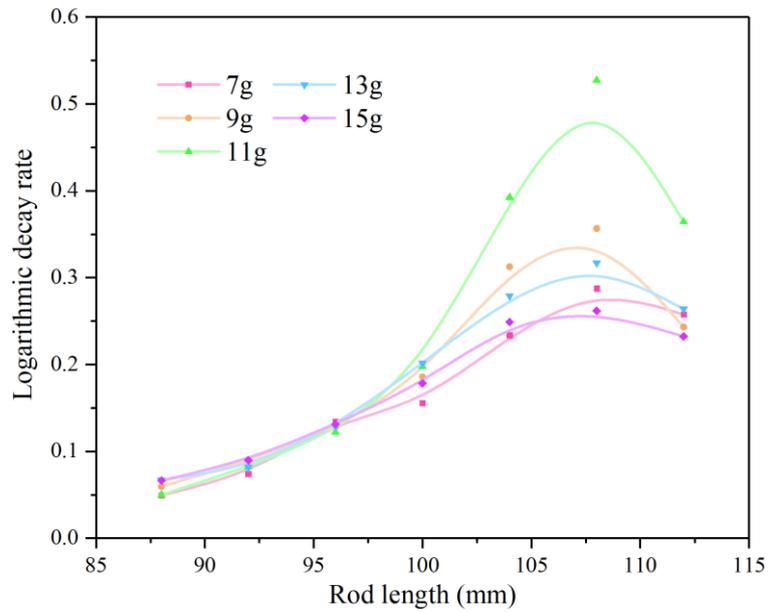

Figure 11. Logarithmic decay rate versus rod length with different magnetic fluid mass.

From the above experiment curves, the SA-MFSA possessed the best damping performance with rod length taken 108mm and MF equal to 11g. The vibration decay time of this optimal SA-MFSA was 9.27s and its vibration attenuation rate was up to 98.7%. Fig. 12 shows displacement responses for initial excitation with the optimal SA-MFSA or equivalent mass, respectively. It is clearly that the vibration decay time in Fig. 12(a) was only a seventieth compared with the decay time in Fig. 12(b).

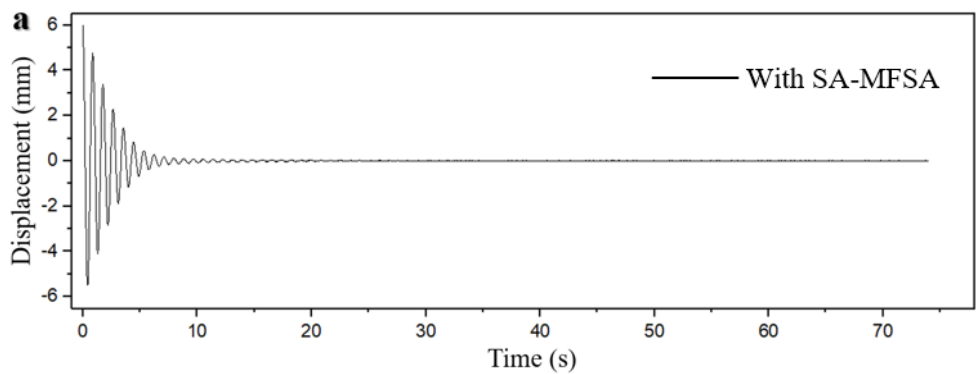

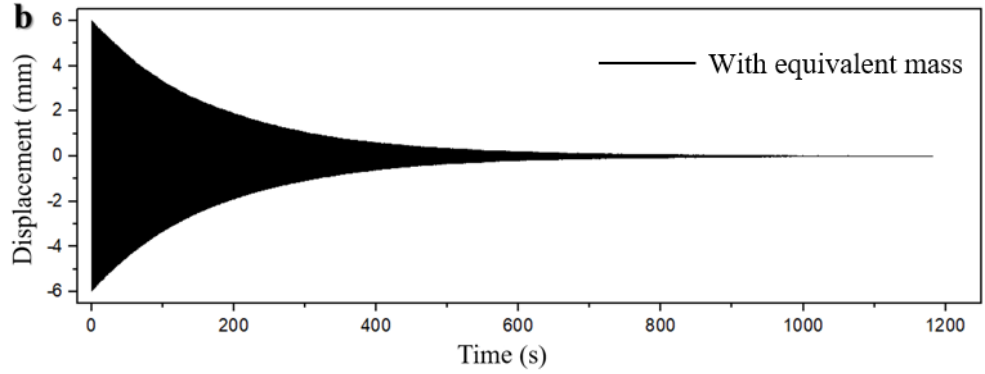

Figure 12. Displacement responses for initial excitation. (a) Response curve with SA-MFSA. (b) Response curve with equivalent mass.

## 6 Conclusion

The vibration problem of spacecraft flexible structure is the key issue obstructing advance in aerospace technology. The MFSA is one of the potential candidates for suppressing such vibration. In order to improve the damping performance of MFSAs, a new SA-MFSA was designed based on the stiffness coefficient formula. The specific conclusions were as follow.

(1) According to the theoretical analysis of stiffness coefficients of other passive dampers, the stiffness coefficient formula which was suitable for MFSAs was got. Besides, the optimal stiffness coefficient was estimated to be equal to 5.25 N/m. Then, this value can guide the design of MFSAs.
(2) A novel SA-MFSA was proposed, which possessed the adjustable stiffness by changing the connecting rod length. On the basis of 3D-printing, a series of SA-MFSAs were fabricated and their stiffness coefficient included the optimal value.
(3) The repulsive force measurement was both simulated and verified by experiments. The simulation and experiment curves were in good agreement. Moreover, the relationship between the repulsive force and offset distance was linear in the small displacement section. It indicated that the rod length corresponding to the optimal stiffness coefficient can be decided by simulation first for subsequence experiments.
(4) The vibration attenuation experiments were carried out and the damping performance of SA-MFSAs was evaluated by the vibration attenuation rate and logarithmic decay rate. As a result, the rod length had a significant influence on the damping efficiency of SA-MFSAs, which meant the stiffness coefficient played an important role in damping performance of SA-MFSAs. It was obvious that SA-MFSA with the optimal stiffness coefficient had the best damping performance.

In summary, the optimal stiffness coefficient can well guide the design of MFSAs. It is hope that the refined stiffness coefficient formula proposed by us can be applied in other liquid dampers in the future.

## 7 Conflict of Interest

The authors declare that the research was conducted in the absence of any commercial or financial relationships that could be construed as a potential conflict of interest.

## 8 Author Contributions

Yanwen Li: Conceptualization, Methodology, Validation, Formal analysis, Investigation, Data Cleansing, Writing, Review, Editing. Decai Li: Funding Acquisition. Yingsong Li: Investigation, Writing.

## 9 Funding

This work was supported by National Natural Science Foundation of China (Grant No. 51735006, 51927810 and U1837206) and Beijing Municipal Natural Science Foundation (Grant No. 3182013).